\documentclass[twocolumn,preprintnumbers,unsortedaddress,amsmath,prb,nobibnotes,altaffilletter]{revtex4}

\usepackage{graphics,graphicx}% Include figure files
\usepackage{dcolumn}% Align table columns on decimal point
\preprint{Physics of Fluids}

\begin{document}

\title{Laboratory experiments for intense vortical structures in turbulence velocity fields}

\author{Hideaki Mouri}
\email{hmouri@mri-jma.go.jp}

\author{Akihiro Hori}
\altaffiliation[Also at ]{Meteorological and Environmental Sensing Technology, Inc., Nanpeidai, Ami 300-0312, Japan}

\author{Yoshihide Kawashima}
\altaffiliation[Also at ]{Meteorological and Environmental Sensing Technology, Inc., Nanpeidai, Ami 300-0312, Japan}
\affiliation{Meteorological Research Institute, Nagamine, Tsukuba 305-0052, Japan}

\date{\today}

\begin{abstract}
Vortical structures of turbulence, i.e., vortex tubes and sheets, are studied using one-dimensional velocity data obtained in laboratory experiments for duct flows and boundary layers at microscale Reynolds numbers from 332 to 1934. We study the mean velocity profile of intense vortical structures. The contribution from vortex tubes is dominant. The radius scales with the Kolmogorov length. The circulation velocity scales with the rms velocity fluctuation. We also study the spatial distribution of intense vortical structures. The distribution is self-similar over small scales and is random over large scales. Since these features are independent of the microscale Reynolds number and of the configuration for turbulence production, they appear to be universal.
\end{abstract}

\maketitle

\section{Introduction}
\label{s1}

Turbulence contains various classes of structures that are embedded in the background random fluctuation. They are important to intermittency as well as mixing and diffusion. Of particular interest are small-scale structures, which could have universal features that are independent of the Reynolds number and of the large-scale flow. We explore such universality using velocity data obtained in laboratory experiments.

We focus on vortical structures, i.e., vortex tubes and sheets. The former is often regarded as the elementary structure of turbulence.\cite{f95,sa97,kida} At low microscale Reynolds numbers, Re$_{\lambda} \alt 200$, direct numerical simulations derived basic parameters of vortex tubes.\cite{kida,vm91,vm94,j93,jw98,tkmsm04} The radii are of the order of the Kolmogorov length $\eta$. The total lengths are of the order of the correlation length $L$. The circulation velocities are of the order of the rms velocity fluctuation $\langle u^2 \rangle ^{1/2}$ or the Kolmogorov velocity $u_K$. Here $\langle \cdot \rangle$ denotes an average. The lifetimes are of the order of the turnover time for energy-containing eddies $L/ \langle u^2 \rangle ^{1/2}$.

For these vortical structures, however, universality has not been established because the behavior at high Reynolds numbers has not been known. At Re$_{\lambda} \agt 200$, a direct numerical simulation is not easy for now.

The promising approach is velocimetry in laboratory experiments. A probe suspended in the flow is used to obtain a one-dimensional cut of the velocity field. The velocity variation is intense at the positions of intense structures. Especially at the positions of intense vortical structures, the variation of the velocity component that is perpendicular to the one-dimensional cut is intense.\cite{p94,mtk99} Thus, the velocity variation offers some information about intense structures, although it is difficult to specify their geometry.

The above approach was taken in several studies.\cite{b96,n97,cg99,mhk03,mhk04} For example, using grid turbulence \cite{mhk03} at Re$_{\lambda} = 105$--329 and boundary layers \cite{mhk04} at Re$_{\lambda} = 295$--1258, we studied the scale, intensity, and spatial distribution of vortical structures. However, the Reynolds number could be increased still more. The dependence of those features on the large-scale flow, i.e., on the configuration for turbulence production, has not been known.

We accordingly use duct flows and boundary layers to compare features of intense vortical structures along the Reynolds number. The highest Reynolds number is Re$_{\lambda} = 1934$, which exceeds those of the prior studies. Our experiments are described in Sec. \ref{s2}. We discuss in Sec. \ref{s3} what information is available from one-dimensional velocity data. The scale, intensity, and spatial distribution of vortical structures are studied in Secs. \ref{s4}--\ref{s6}. The conclusions are summarized in Sec. \ref{s7}.

\begingroup
%\squeezetable
\begin{table*}
\caption{\label{t1} 
Experimental conditions and turbulence parameters: duct-exit or incoming-flow velocity $U_{\ast}$, coordinates $x$ and $z$ of the measurement position, mean streamwise velocity $U$, sampling frequency $f_s$, kinematic viscosity $\nu$, mean energy dissipation rate $\langle \varepsilon \rangle = 15 \nu \langle (\partial _x v)^2 \rangle /2$, rms velocity fluctuations $\langle u^2 \rangle ^{1/2}$ and $\langle v^2 \rangle ^{1/2}$, Kolmogorov velocity $u_K = ( \nu \langle \varepsilon \rangle )^{1/4}$, rms spanwise-velocity increment over the sampling interval $\langle \delta v_s^2 \rangle ^{1/2} = \langle [v(x+U/2f_s)-v(x-U/2f_s)]^2 \rangle^{1/2}$, correlation lengths $L_u = \int^{\infty}_{0} \langle u(x+r) u(x) \rangle / \langle u^2 \rangle dr$ and $L_v = \int ^{\infty}_{0} \langle v(x+r) v(x) \rangle / \langle v^2 \rangle dr$, Taylor microscale $\lambda = [2 \langle v^2 \rangle / \langle (\partial _x v)^2 \rangle ]^{1/2}$, Kolmogorov length $\eta = (\nu ^3 / \langle \varepsilon \rangle )^{1/4}$, and microscale Reynolds number Re$_{\lambda} = \lambda \langle v^2 \rangle ^{1/2} / \nu$. The velocity derivative was obtained as $\partial _x v = [ 8v(x+r)-8v(x-r)-v(x+2r)+v(x-2r) ] / 12r$ with $r=U/f_s$.}

\begin{ruledtabular}
\begin{tabular}{llccccccccccc}
                              &                 & \multicolumn{5}{c}{Duct flow}               & \multicolumn{6}{c}{Boundary layer}         \\
\cline{3-7}
\cline{8-13}
                              & Units           & 1      & 2      & 3      & 4      & 5       & 6      & 7      & 8      & 9      & 10     & 11     \\ 
\hline
$U_{\ast}$                    & m s$^{-1}$      & 11     & 23     & 34     & 45     & 55      & 2      & 4      & 8      & 12     & 16     & 20     \\
$x$                           & m               & 15.5   & 15.5   & 15.5   & 15.5   & 15.5    & 12.5   & 12.5   & 12.5   & 12.5   & 12.5   & 12.5    \\
$z$                           & m               & 0.60   & 0.60   & 0.60   & 0.60   & 0.60    & 0.35   & 0.35   & 0.30   & 0.30   & 0.25   & 0.25      \\
$U$                           & m s$^{-1}$      & 4.26   & 8.66   & 13.0   & 17.3   & 21.2    & 1.51   & 3.12   & 5.84   & 8.75   & 10.8   & 13.6   \\
$f_s$                         & kHz             & 12     & 34     & 60     & 80     & 100     & 4      & 10     & 22     & 38     & 60     & 74     \\
$\nu$                         & cm$^2$ s$^{-1}$ & 0.141  & 0.141  & 0.142  & 0.142  & 0.142   & 0.141  & 0.142  & 0.142  & 0.142  & 0.142  & 0.143  \\
$\langle \varepsilon \rangle$ & m$^2$ s$^{-3}$  & 0.405  & 2.87   & 9.13   & 19.2   & 34.4    & 0.0332 & 0.226  & 1.85   & 5.37   & 13.8   & 24.6   \\ 
$\langle u^2 \rangle ^{1/2}$  & m s$^{-1}$      & 0.694  & 1.38   & 2.11   & 2.84   & 3.46    & 0.283  & 0.582  & 1.18   & 1.80   & 2.46   & 3.14   \\
$\langle v^2 \rangle ^{1/2}$  & m s$^{-1}$      & 0.666  & 1.34   & 2.04   & 2.69   & 3.32    & 0.242  & 0.475  & 0.973  & 1.46   & 1.98   & 2.51   \\
$u_K$                         & m s$^{-1}$      & 0.0489 & 0.0798 & 0.107  & 0.128  & 0.149   & 0.0262 & 0.0423 & 0.0716 & 0.0934 & 0.118  & 0.137  \\
$\langle \delta v_s^2\rangle^{1/2}$& m s$^{-1}$ & 0.0219 & 0.0421 & 0.0637 & 0.0918 & 0.121   & 0.00667& 0.0143 & 0.0349 & 0.0517 & 0.0646 & 0.0882  \\
$L_u$                         & cm              & 55.4   & 55.8   & 51.9   & 46.3   & 42.2    & 43.9   & 48.5   & 41.5   & 43.4   & 47.3   & 44.0   \\
$L_v$                         & cm              & 14.9   & 14.5   & 14.7   & 14.2   & 14.2    & 6.67   & 6.87   & 6.33   & 5.83   & 5.71   & 6.11   \\
$\lambda$                     & cm              & 1.52   & 1.15   & 0.986  & 0.895  & 0.826   & 1.93   & 1.46   & 1.04   & 0.919  & 0.776  & 0.742  \\
$\eta$                        & cm              & 0.0288 & 0.0177 & 0.0133 & 0.0111 & 0.00955 & 0.0539 & 0.0335 & 0.0198 & 0.0152 & 0.0120 & 0.0104 \\
Re$_{\lambda}$                &                 & 719    & 1098   & 1416   & 1693   & 1934    & 332    & 488    & 716    & 945    & 1080   & 1304   \\

\end{tabular}
\end{ruledtabular}
\end{table*}
\endgroup

\section{Experiment}
\label{s2}

\subsection{Duct flow}

The experiment was done in a wind tunnel of the Meteorological Research Institute. Table \ref{t1} lists experimental conditions and turbulence parameters. The microscale Reynolds number ranges from Re$_{\lambda} = 719$ to 1934.

We use coordinates $x$, $y$, and $z$ in the streamwise, spanwise, and floor-normal directions. The corresponding flow velocities are $U+u$, $v$, and $w$. Here $U$ is the mean while $u$, $v$, and $w$ are the fluctuations. The origin $x = y = z = 0$\,m is on the tunnel floor at the entrance to the test section. Its size was $\delta x = 18$\,m, $\delta y = 3$\,m, and $\delta z = 2$\,m. We placed a rectangular duct with width $\delta y = 1.3$\,m and $\delta z = 1.4$\,m at $x = -2$\,m. The duct center was on the tunnel axis.

For the flow velocities $U_{\ast}$ from 11 to 55\,m\,s$^{-1}$ at the duct exit, we measured the streamwise and spanwise velocities at $x = 15.5$\,m and $z = 0.6$\,m. The flow was turbulent there. During each of the measurements, the flow temperature was constant within $\pm 1^{\circ}$C.

We used a hot-wire anemometer, which was composed of a constant-temperature system and a crossed-wire probe. The wires were made of tungsten, 5\,$\mu$m in diameter, 1.25\,mm in sensing length, 1.4\,mm in separation, and oriented at $\pm 45 ^{\circ}$ to the streamwise direction.

The signal was linearized, low-pass filtered at 24\,dB per octave, and then digitally sampled at 16-bit resolution. We set the sampling frequency $f_s$ as high as possible, on condition that high-frequency noise was not significant in the power spectrum. The filter cutoff frequency was one-half of the sampling frequency. For $f_s \le 50$\,kHz, we obtained $10^8$ data of the streamwise and spanwise velocities. For $f_s > 50$\,kHz, we obtained $4 \times 10^8$ data of the spanwise velocity alone. This is due to a $f_s$ limit of our sampling device. Supplementarily, we obtained $4 \times 10^7$ data of the streamwise and spanwise velocities at $f_s = 50$\,kHz.

The temporal variations were converted into the spatial variations using Taylor's frozen-eddy hypothesis, which requires that the turbulence strength $\langle u^2 \rangle ^{1/2}/U$ is small enough. This requirement is satisfied in our experiments where $\langle u^2 \rangle ^{1/2}/U \lesssim 0.2$. Even for $\langle u^2 \rangle ^{1/2}/U \simeq 0.3$, the validity of Taylor's hypothesis was confirmed using data obtained simultaneously with two probes separated by streamwise distances.\cite{sd98}

By assuming local isotropy $\langle ( \partial_x v )^2 \rangle = 2 \langle ( \partial_x u )^2 \rangle$, the small-scale statistics were obtained from the spanwise-velocity derivative $\partial_x v$, instead of the usual streamwise-velocity derivative $\partial_x u $. At small scales, the $u$ component measured by a crossed-wire probe is contaminated with the $w$ component that is perpendicular to the two wires of the probe.\cite{note0} The $v$ component is free from such contamination.

\subsection{Boundary layer}

The experiment was done in the same wind tunnel with the same instruments as for the duct flows. Table \ref{t1} lists experimental conditions and turbulence parameters. They are close to those in our prior study,\cite{mhk04} except that the sampling frequency is higher and the data are longer. The microscale Reynolds number ranges from Re$_{\lambda} = 332$ to 1304.

Over the entire floor of the test section of the wind tunnel, we placed blocks as roughness for the boundary layer. The block size was $\delta x = 0.06$\,m, $\delta y = 0.21$\,m, and $\delta z = 0.11$\,m. The spacing of adjacent blocks was $\delta x = \delta y = 0.5$\,m. 

For the incoming-flow velocities $U_{\ast}$ from 2 to 20\,m\,s$^{-1}$, we measured the streamwise and spanwise velocities at $x = 12.5$\,m.  The boundary layer was well developed there. While the 99\% thickness was 0.8\,m, the displacement thickness was 0.2\,m.\cite{mhk04} The measurement height was $z = 0.25$--0.35\,m in the log-law sublayer.

\begin{figure}[t]
\resizebox{5cm}{!}{\includegraphics*[3cm,6cm][20cm,23cm]{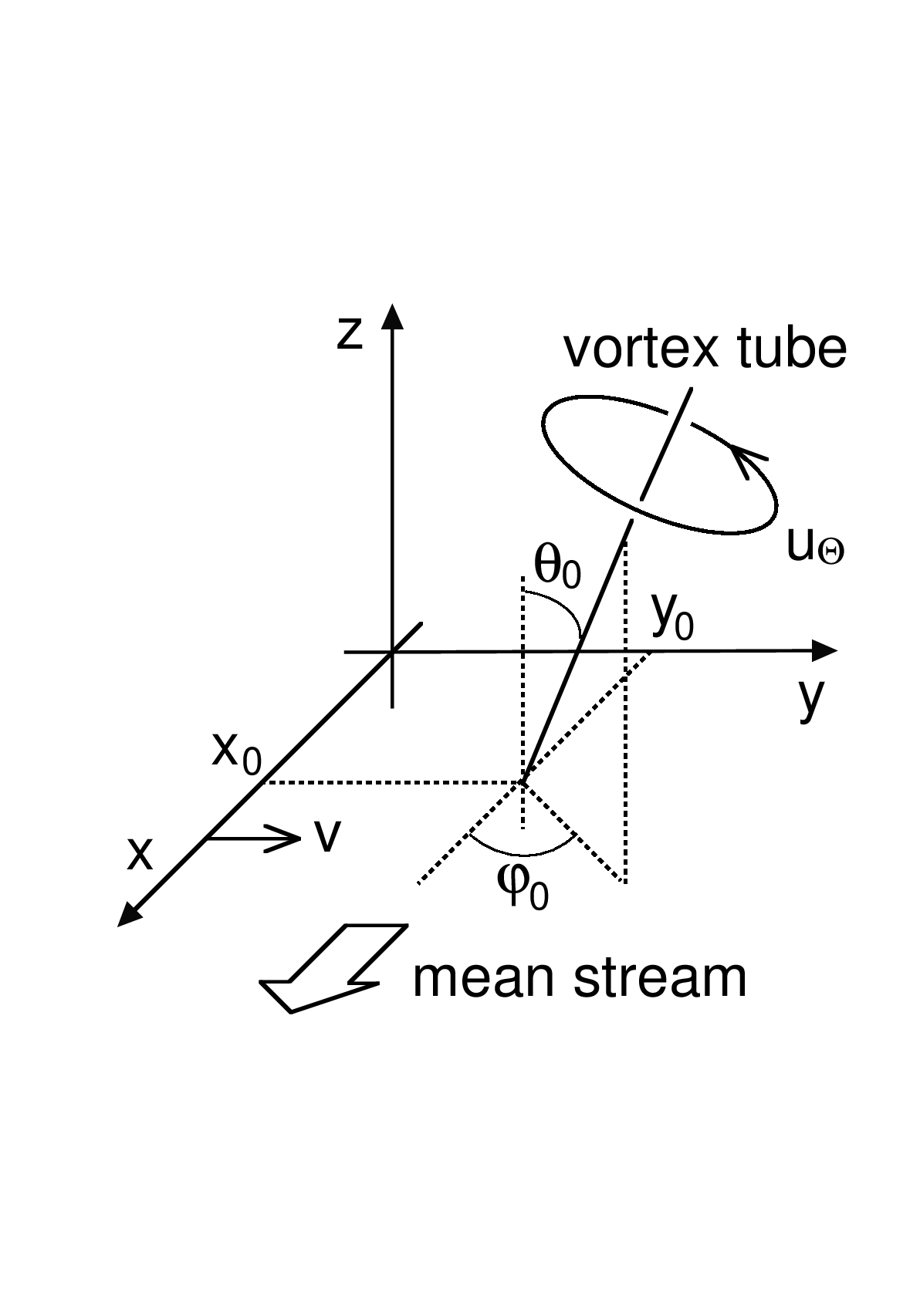}}
\caption{\label{f1} Sketch of a vortex tube penetrating the $(x,y)$ plane at a point $(x_0,y_0)$. The inclination is $(\theta_0, \varphi_0)$. The circulation velocity is $u_{\Theta}$. We consider the spanwise velocity $v$ along the $x$ axis in the mean stream direction.}
\end{figure}

\section{Burgers Vortex}
\label{s3}

By using the Burgers vortex, an idealized model for vortex tubes, we discuss what information is available from a one-dimensional cut of the velocity field. The Burgers vortex is an axisymmetric steady circulation in a strain field. In cylindrical coordinates, the circulation $u_{\Theta}$ and strain field $(u_R, u_Z)$ are
%%%%%%%%%%%%%%%%%%%%%%%%%%%%%%%%%%%%%%%%%%%%%%%%%%%%%%%%%%%%%%%%%%%%%%%%%
\begin{subequations}
\begin{eqnarray}
& &u_{\Theta} \propto \frac{\nu}{a_0 R} \left[ 1 - \exp \left( - \frac{a_0 R^2}{4 \nu} \right) \right],
\\
& &\left( u_R, u_Z \right) = \left(- \frac{a_0 R}{2}, a_0Z \right) .
\end{eqnarray}
\end{subequations}
%%%%%%%%%%%%%%%%%%%%%%%%%%%%%%%%%%%%%%%%%%%%%%%%%%%%%%%%%%%%%%%%%%%%%%%%%
Here $\nu$ is the kinematic viscosity and $a_0$ $(> 0)$ is a constant. The circulation is maximal at $R$ = $R_0$ = $2.24 (\nu / a_0)^{1/2}$. We regard $R_0$ as the vortex radius.

There are other models for vortex tubes, e.g., the Lundgren spirals.\cite{l82} We use the Burgers vortex alone because detailed information about individual structures is anyway not available from one-dimensional velocity data. 

\begin{figure}[t]
\resizebox{7.5cm}{!}{\includegraphics*[4cm,16cm][17.5cm,26cm]{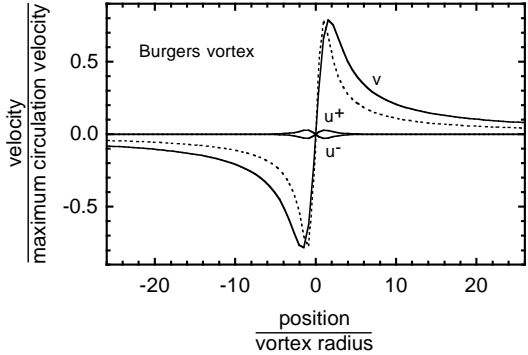}}
\caption{\label{f2} Mean profiles in the streamwise ($u$) and spanwise ($v$) velocities for the Burgers vortices with random positions $(x_0, y_0)$ and inclinations $(\theta_0,\varphi_0)$. The $u$ profile is separately shown for $\partial _x u > 0$ ($u^+$) and $\partial _x u \le 0$ ($u^-$) at $x = 0$. The position $x$ and velocities are normalized by the radius and maximum circulation velocity of the Burgers vortices. The dotted line is the $v$ profile of the Burgers vortex for $x_0 = y_0 = \theta_0 = 0$, the peak value of which is scaled to that of the mean $v$ profile.}
\end{figure}

Suppose that the axis of the Burgers vortex penetrates the $(x,y)$ plane at a point $(x_0,y_0)$ as shown in Fig. \ref{f1}. The $x$ and $y$ axes are in the streamwise and spanwise directions. If the inclination of the Burgers vortex is $(\theta_0, \varphi_0)$ in spherical coordinates, the streamwise ($u$) and spanwise ($v$) components of the circulation flow $u_{\Theta}$ along the $x$ axis are
%%%%%%%%%%%%%%%%%%%%%%%%%%%%%%%%%%%%%%%%%%%%%%%%%%%%%%%%%%%%%%%%%%%%%%%%%
\begin{subequations}
\begin{eqnarray}
\label{eq2a} & &u(x-x_0) = \frac{y_0     \cos \theta_0}{R}u_{\Theta} (R) , \\
\label{eq2b} & &v(x-x_0) = \frac{(x-x_0) \cos \theta_0}{R}u_{\Theta} (R),
\end{eqnarray}
\end{subequations}
%%%%%%%%%%%%%%%%%%%%%%%%%%%%%%%%%%%%%%%%%%%%%%%%%%%%%%%%%%%%%%%%%%%%%%%%%
with
%%%%%%%%%%%%%%%%%%%%%%%%%%%%%%%%%%%%%%%%%%%%%%%%%%%%%%%%%%%%%%%%%%%%%%%%%
\begin{eqnarray}
\label{eq3}
R^2 &=& (x-x_0)^2       ( 1 - \sin ^2 \theta_0 \cos ^2 \varphi_0 )  \\ \nonumber
    &+&  y_0 ^2          ( 1 - \sin ^2 \theta_0 \sin ^2 \varphi_0 ) \\
    &+&  2(x-x_0) y_0 \sin ^2 \theta_0 \sin \varphi_0 \cos \varphi_0. \nonumber
\end{eqnarray}
%%%%%%%%%%%%%%%%%%%%%%%%%%%%%%%%%%%%%%%%%%%%%%%%%%%%%%%%%%%%%%%%%%%%%%%%%
Those of the radial inflow $u_R$ of the strain field are
%%%%%%%%%%%%%%%%%%%%%%%%%%%%%%%%%%%%%%%%%%%%%%%%%%%%%%%%%%%%%%%%%%%%%%%%%
\begin{widetext}
\begin{subequations}
\begin{eqnarray}
\label{eq4a}
& &u(x-x_0) =   \frac{(x-x_0) ( 1 - \sin ^2 \theta_0 \cos ^2 \varphi_0 )   + y_0 \sin ^2 \theta_0 \sin \varphi_0 \cos \varphi_0 }{R}u_R (R) , \\
\label{eq4b}
& &v(x-x_0) = - \frac{(x-x_0) \sin ^2 \theta_0 \sin \varphi_0 \cos \varphi_0 + y_0 ( 1 - \sin ^2 \theta_0 \sin ^2 \varphi_0 )   }{R}u_R (R).
\end{eqnarray}
\end{subequations}
\end{widetext}
%%%%%%%%%%%%%%%%%%%%%%%%%%%%%%%%%%%%%%%%%%%%%%%%%%%%%%%%%%%%%%%%%%%%%%%%%
When the Burgers vortex is close to the $x$ axis and is not heavily inclined, i.e., $|y_0| \alt R_0$ and $\theta_0 \simeq 0$, the spanwise velocity is dominated by the small-scale circulation flow [Eq. (\ref{eq2b})]. The streamwise velocity is dominated by the large-scale radial inflow [Eq. (\ref{eq4a})]. When $|y_0| \gg R_0$ or $\theta_0 \gg 0$, the signal is weak at least for the small-scale variation in the spanwise velocity.

Then, suppose that velocity data are obtained on a one-dimensional cut of a flow that consists of vortex tubes and the background random fluctuation. The tube position $(x_0,y_0)$ and inclination $(\theta_0,\varphi_0)$ are supposed to be random.  This is likely in our experiments, where turbulence was almost isotropic because the measured ratio $\langle u^2 \rangle / \langle v^2 \rangle$ is not far from unity (Table \ref{t1}). The vortex tubes induce small-scale variations in the spanwise velocity. If we consider intense velocity variations above a high threshold, their scale and amplitude are close to the radius and circulation velocity of intense vortex tubes with $|y_0| \alt R_0$ and $\theta_0 \simeq 0$. To demonstrate this, mean profiles are calculated for the circulation flows $u _{\Theta}$ of the Burgers vortices with random positions $(x_0, y_0)$ and inclinations $(\theta_0,\varphi_0)$. Their radii $R_0$ and maximum circulation velocities $V_0 = u_{\Theta}(R_0)$ are set to be the same. We consider the Burgers vortices with $| \partial _x v |$ at $x=0$ being above a threshold, $| \partial _x v |/3$ at $x=0$ for $x_0 = y_0 = \theta_0 = 0$. When $\partial _x v$ is negative, the sign of the $v$ signal is inverted before the averaging. The result is shown in Fig. \ref{f2}. Despite the relatively low threshold, the scale and peak amplitude of the mean $v$ profile are still close to those of the $v$ profile for $x_0 = y_0 = \theta_0 = 0$ (dotted line). The extended tails are due to the Burgers vortices with $|y_0| \gg R_0$ or $\theta_0 \gg 0$.

\begin{figure}[t]
\resizebox{7.5cm}{!}{\includegraphics*[4cm,9.5cm][17.5cm,26cm]{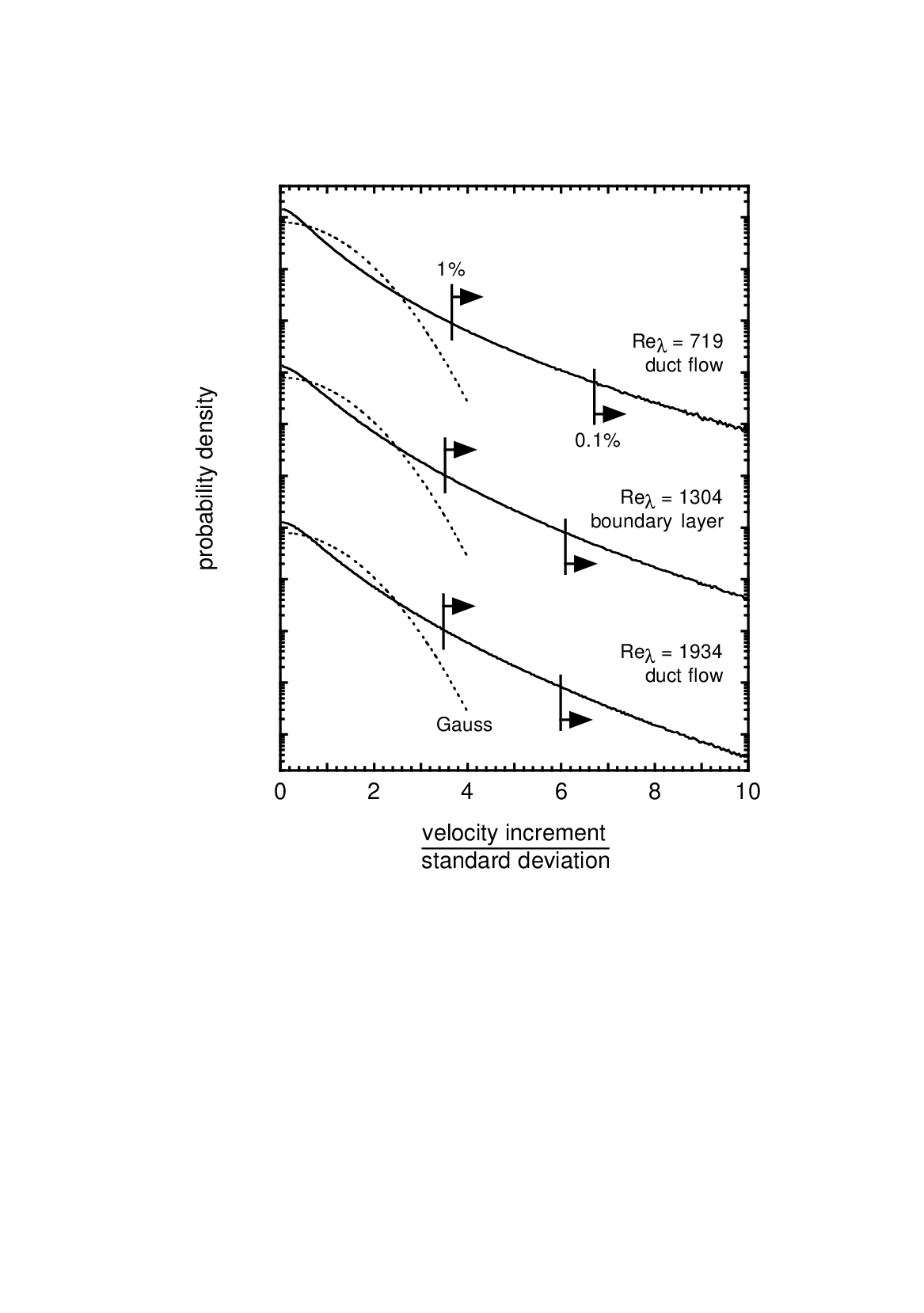}}
\caption{\label{f3} Probability density distribution of the absolute spanwise-velocity increment $\vert v(x+U/2f_s) - v(x-U/2f_s) \vert$ at Re$_{\lambda} = 719$, 1304, and 1934. The distribution is vertically shifted by a factor $10^3$. The increment is normalized by $\langle \delta v_s^2\rangle^{1/2} = \langle [v(x+U/2f_s) - v(x-U/2f_s)]^2 \rangle ^{1/2}$. The arrows indicate the ranges for intense vortical structures, which share 0.1 and 1\% of the total. The dotted line denotes the Gaussian distribution.}
\end{figure} 

\begin{figure}[t]
\resizebox{7.5cm}{!}{\includegraphics*[4cm,9.5cm][17.5cm,26cm]{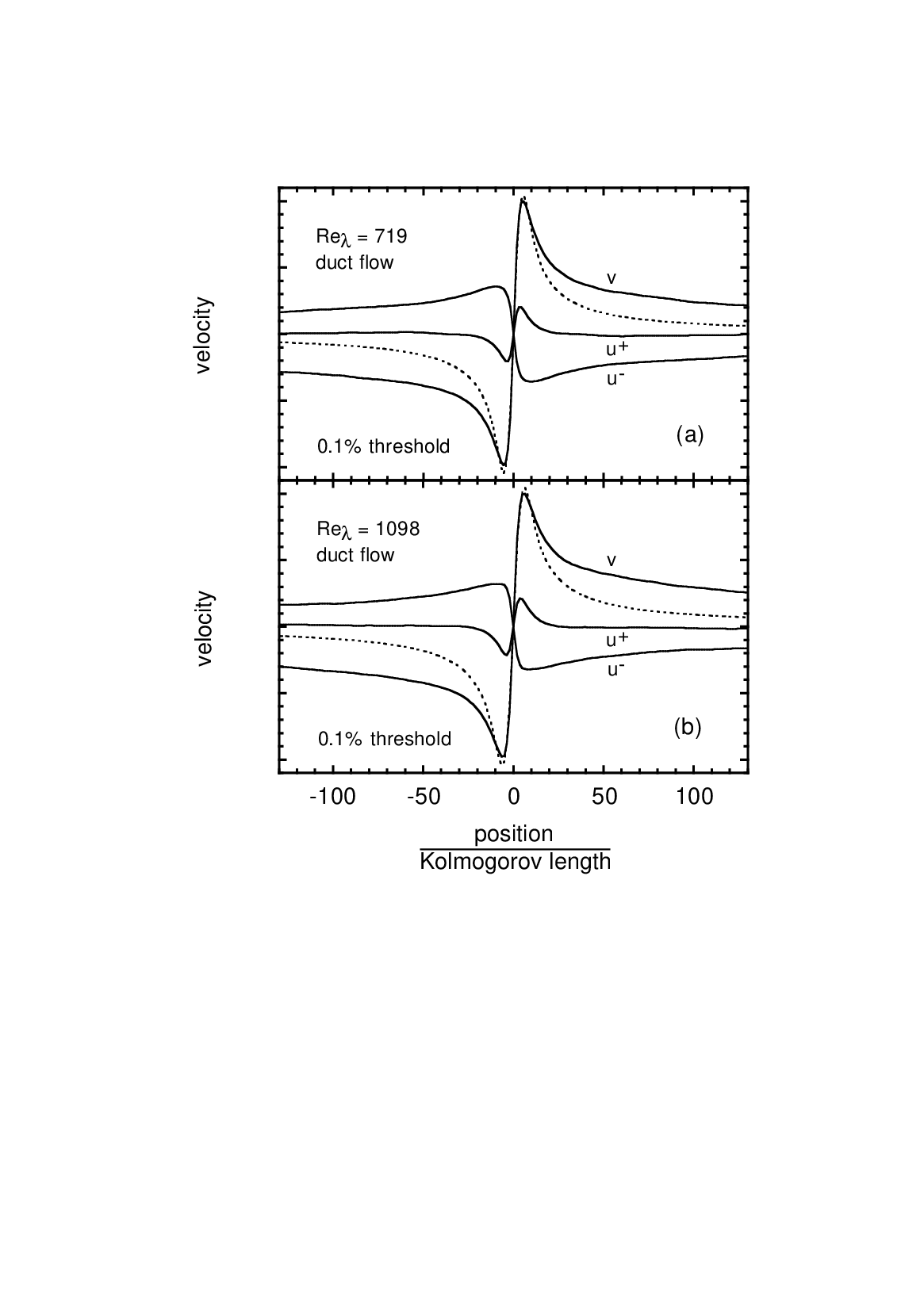}}
\caption{\label{f4} Mean profiles of intense vortical structures for the 0.1\% threshold in the streamwise ($u$) and spanwise ($v$) velocities. (a)  Re$_{\lambda} = 719$. (b) Re$_{\lambda} = 1098$. The $u$ profile is separately shown for $\partial _x u > 0$ ($u^+$) and $\partial _x u \le 0$ ($u^-$) at $x = 0$. The position $x$ is normalized by the Kolmogorov length. The velocities are normalized by the peak value of the $v$ profile. We also show the $v$ profile of the Burgers vortex for $x_0 = y_0 = \theta_0 = 0$ by a dotted line.}
\end{figure}

\section{Mean Velocity Profile}
\label{s4}

Mean profiles of intense vortical structures in the streamwise ($u$) and spanwise ($v$) velocities are extracted, by averaging signals centered at the position where the absolute spanwise-velocity increment $\vert v(x+r/2) - v(x-r/2) \vert$ is above a threshold.\cite{mtk99,cg99,mhk03,mhk04,note2} The scale $r$ is the sampling interval $U/f_s$. The threshold is such that 0.1\% or 1\% of the increments are used for the averaging (hereafter, the 0.1\% or 1\% threshold). These increments comprise the tail of the probability density distribution of all the increments as in Fig. \ref{f3}.\cite{note1} Example of the results are shown in Fig. \ref{f4}.\cite{note1}

The $v$ profile in Fig. \ref{f4} is close to the $v$ profile in Fig. \ref{f2}. Hence, the contribution from vortex tubes is dominant. The contribution from vortex sheets is not dominant. If it were dominant, the $v$ profile should exhibit some kind of step.\cite{n97} Direct numerical simulations at Re$_{\lambda} \alt 200$ revealed that intense vorticity tends to be organized into tubes rather than sheets.\cite{vm91,vm94,j93,jw98,mj04,mj05} This tendency appears to exist up to Re$_{\lambda} \simeq 2000$. Vortex sheets might contribute to the extended tails in Fig. \ref{f4}. They are more pronounced than those in Fig. \ref{f2}. Here it should be noted that our discussion is somewhat simplified because there is no strict division between vortex tubes and sheets in real turbulence.

\begingroup
\squeezetable
\begin{table*}
\caption{\label{t2} 
Parameters for intense vortical structures: radius $R_0$, maximum circulation velocity $V_0$, Reynolds number Re$_0 = R_0 V_0 / \nu$ and small-scale clustering exponent $\mu_0$. We also list the threshold level $\tau_0$.}

\begin{ruledtabular}
\begin{tabular}{llccccccccccc}
        &                                      & \multicolumn{5}{c}{Duct flow}                 & \multicolumn{6}{c}{Boundary layer}         \\
\cline{3-7}
\cline{8-13}
         & Units                               & 1     & 2     & 3     & 4     & 5     & 6     & 7     & 8     & 9     & 10    & 11    \\ 
\hline
\multicolumn{13}{c}{For the 0.1\% threshold} \\
$\tau_0$ & $\langle \delta v_s^2\rangle^{1/2}$ & 6.69  & 6.50  & 6.28  & 6.17  & 6.01  & 6.45  & 6.54  & 6.43  & 6.29  & 6.25  & 6.12  \\
$R_0$    & $\eta$                              & 5.21  & 5.64  & 5.47  & 5.87  & 6.20  & 5.18  & 5.24  & 5.56  & 5.97  & 5.66  & 6.15  \\
$V_0$    & $\langle v^2 \rangle ^{1/2}$        & 0.637 & 0.595 & 0.593 & 0.610 & 0.629 & 0.822 & 0.745 & 0.701 & 0.677 & 0.700 & 0.700 \\
$V_0$    & $u_K$                               & 8.67  & 10.0  & 11.3  & 12.8  & 14.0  & 7.60  & 8.37  & 9.53  & 10.6  & 11.7  & 12.8  \\
Re$_0$   &                                     & 45.2  & 56.4  & 61.8  & 75.1  & 86.8  & 39.4  & 43.9  & 53.0  & 63.3  & 66.2  & 78.7  \\
$\mu_0$  &                                     & 0.867 & 0.879 & 0.896 & 0.919 & 0.907 & 0.840 & 0.855 & 0.921 & 0.924 & 0.995 & 0.992 \\
\multicolumn{13}{c}{For the 1\% threshold} \\
$\tau_0$ & $\langle \delta v_s^2\rangle^{1/2}$ & 3.66  & 3.62  & 3.57  & 3.55  & 3.52  & 3.61  & 3.64  & 3.61  & 3.57  & 3.57  & 3.53  \\
$R_0$    & $\eta$                              & 6.50  & 6.61  & 6.92  & 6.76  & 7.03  & 6.53  & 6.44  & 6.92  & 7.38  & 7.14  & 7.19  \\
$V_0$    & $\langle v^2 \rangle ^{1/2}$        & 0.449 & 0.406 & 0.395 & 0.401 & 0.411 & 0.607 & 0.533 & 0.489 & 0.466 & 0.468 & 0.465 \\
$V_0$    & $u_K$                               & 6.11  & 6.83  & 7.53  & 8.41  & 9.17  & 5.61  & 5.98  & 6.65  & 7.28  & 7.83  & 8.53  \\
Re$_0$   &                                     & 39.7  & 45.2  & 52.1  & 56.9  & 64.5  & 36.6  & 38.5  & 46.0  & 53.7  & 55.9  & 61.3  \\
$\mu_0$  &                                     & 1.02  & 1.06  & 1.07  & 1.09  & 1.07  & 0.884 & 0.953 & 0.998 & 1.03  & 1.08  & 1.09  \\
\end{tabular}
\end{ruledtabular}
\end{table*}
\endgroup

By fitting the $v$ profile in Fig. \ref{f4} around its peaks by the $v$ profile of the Burgers vortex for $x_0 = y_0 = \theta_0 = 0$ (dotted line), we estimate the radius $R_0$ and maximum circulation velocity $V_0$. The measured velocity $v_m$ is considered to be the true velocity $v_t$ averaged over the probe size in the streamwise direction, $\delta x_p = 1$\,mm:
%%%%%%%%%%%%%%%%%%%%%%%%%%%%%%%%%%%%%%%%%%%%%%%%%%%%%%%%%%%%%%%%%%%%%%%%%
\begin{equation}
\label{eq5}
v_m(x) = \frac{1}{\delta x_p} \int^{\delta x_p/2}_{-\delta x_p/2} v_t(x+r) dr.
\end{equation} 
%%%%%%%%%%%%%%%%%%%%%%%%%%%%%%%%%%%%%%%%%%%%%%%%%%%%%%%%%%%%%%%%%%%%%%%%%
For all the data, the $R_0$ and $V_0$ values are summarized in Table \ref{t2}. They characterize the scale and intensity of vortical structures, even if they are not the Burgers vortices. The radius $R_0$ is several times the Kolmogorov length $\eta$. The maximum circulation velocity $V_0$ is several tenths of the rms velocity fluctuation $\langle v^2 \rangle ^{1/2}$ and several times the Kolmogorov velocity $u_K$. Similar results were obtained from direct numerical simulations\cite{kida,vm91,j93,jw98,tkmsm04} and laboratory experiments\cite{b96,n97,mhk03,mhk04} at the lower Reynolds numbers, Re$_{\lambda} \alt 1300$.

The $u$ profile in Fig. \ref{f4} is separated for $\partial _x u > 0$ ($u^+$) and $\partial _x u \le 0$ ($u^-$) at $x = 0$. Since the contamination with the $w$ component\cite{note0} induces a symmetric positive excursion,\cite{mhk03,sm05,note3} we decomposed the $u^{\pm}$ profiles into symmetric and antisymmetric components and show only the antisymmetric components.\cite{mhk04} The $u^{\pm}$ profiles in Fig. \ref{f4} have larger amplitudes than those in Fig. \ref{f2}. Hence, the $u^{\pm}$ profiles in Fig. \ref{f4} are dominated by the circulation flows $u_{\Theta}$ of vortex tubes that passed the probe with some incidence angles to the mean flow direction,\cite{b96} $\tan^{-1}[v/(U+u)]$.@The radial inflow $u_R$ of the strain field is not discernible, except that the $u^-$ profile has a larger amplitude than the $u^+$ profile.\cite{mhk03,mhk04} Unlike the Burgers vortex, a real vortex tube is not always oriented to the stretching direction.\cite{vm91,vm94,j93,jw98,tkmsm04,t92}

\begin{figure}[b]
\resizebox{7.5cm}{!}{\includegraphics*[4cm,9.5cm][17.5cm,26cm]{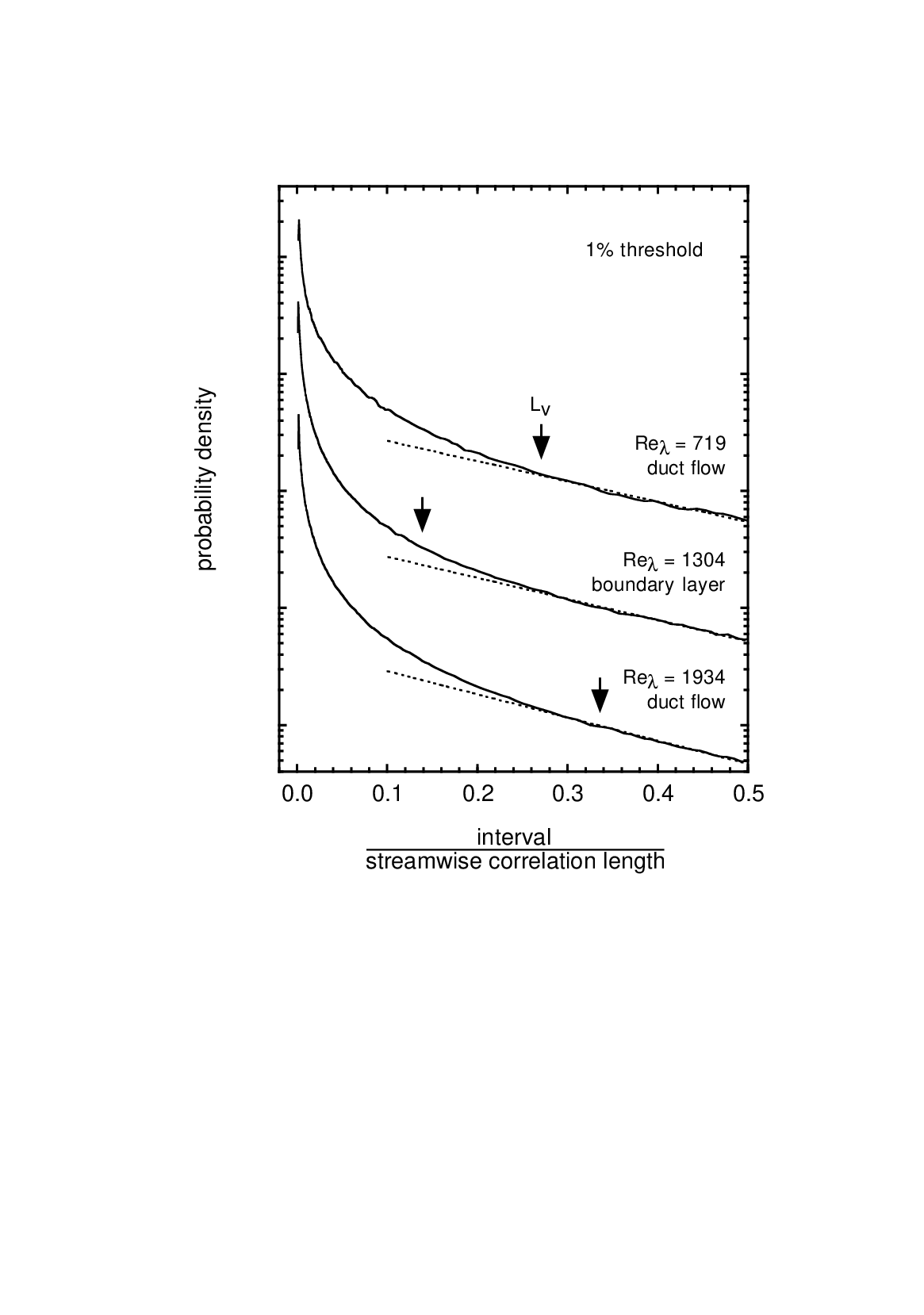}}
\caption{\label{f5} Probability density distribution of interval between intense vortical structures for the 1\% threshold at Re$_{\lambda} = 719$, 1304, and 1934. The distribution is normalized by the amplitude of the exponential tail (dotted line), and it is vertically shifted by a factor 10. The interval is normalized by the streamwise correlation length $L_u$.  The arrow indicates the spanwise correlation length $L_v$.}
\end{figure} 

\begin{figure}[t]
\resizebox{7.5cm}{!}{\includegraphics*[4cm,9.5cm][17.5cm,26cm]{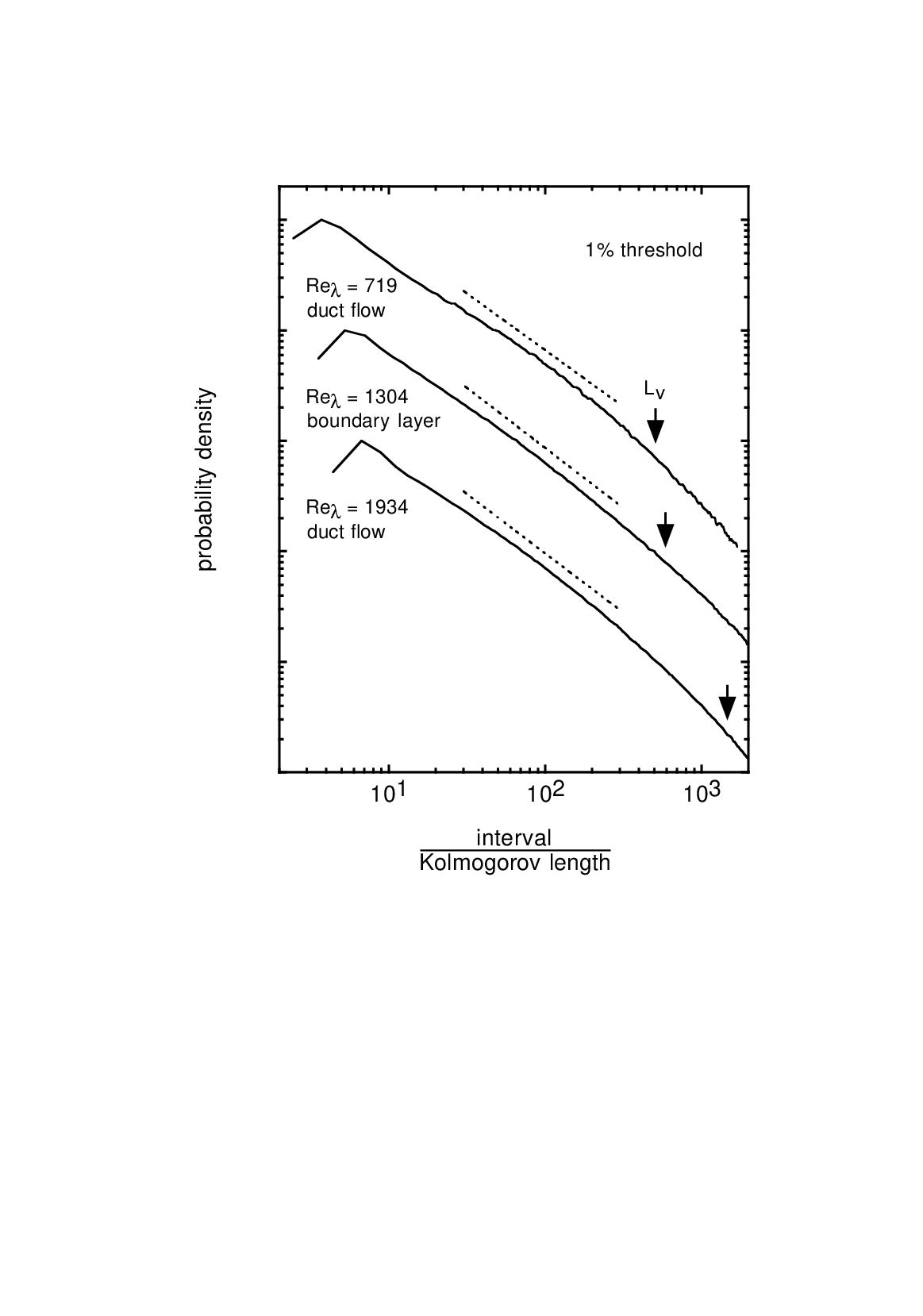}}
\caption{\label{f6} Probability density distribution of interval between intense vortical structures for the 1\% threshold at Re$_{\lambda} = 719$, 1304, and 1934. The distribution is normalized by the peak value, and it is vertically shifted by a factor 10. The dotted line indicates the power-law slope from $30\eta$ to $300\eta$. The interval is normalized by the Kolmogorov length $\eta$. The arrow indicates the spanwise correlation length $L_v$.}
\end{figure}

\section{Spatial Distribution}
\label{s5}

The spatial distribution of intense vortical structures is studied using the distribution of interval $\delta x_0$ between successive intense velocity increments.\cite{cg99,mhk03,mhk04,mj05} The intense velocity increment is defined in the same manner as for the mean velocity profiles in Sec. \ref{s4}. Since they are dominated by vortex tubes, we expect that the distribution of intense vortical structures studied here is also essentially the distribution of intense vortex tubes. Examples of the probability density distribution $P(\delta x_0)$ are shown in Figs. \ref{f5} and \ref{f6}.\cite{note1}

The probability density distribution has an exponential tail\cite{mhk03,mhk04} that appears linear on the semi-log plot of Fig. \ref{f5}. This exponential law is characteristic of intervals for a Poisson process of random and independent events. The large-scale distribution of intense vortical structures is random and independent.

Below the spanwise correlation length $L_v$, the probability density is enhanced over that for the exponential distribution.\cite{mhk04} Thus,  intense vortical structures cluster together below the energy-containing scale. In fact, direct numerical simulations revealed that intense vortex tubes lie on borders of energy-containing eddies.\cite{j93}

Over small intervals, the probability density distribution  is a power law\cite{cg99,mj05} that appears linear on the log-log plot of Fig. \ref{f6}:
%%%%%%%%%%%%%%%%%%%%%%%%%%%%%%%%%%%%%%%%%%%%%%%%%%%%%%%%%%%%%%%%%%%%%%%%%
\begin{equation} 
\label{eq6}
P(\delta x_0) \propto \delta x_0 ^{- \mu_0}.
\end{equation}
%%%%%%%%%%%%%%%%%%%%%%%%%%%%%%%%%%%%%%%%%%%%%%%%%%%%%%%%%%%%%%%%%%%%%%%%%
Thus, the small-scale clustering of intense vortical structures is self-similar and has no characteristic scale.\cite{mj05} Table \ref{t2} lists the clustering exponent $\mu_0$ estimated over intervals from $\delta x_0 = 30 \eta$ to $300 \eta$. Its value is close to unity.

The exponential law over large intervals and the power law over small intervals were also found in laboratory experiments for regions of low pressure.\cite{affl94,cdc95,vsg95,lvmb00} They are associated with vortex tubes, although their radii tend to be larger than those of intense vortical structures studied here.\cite{lvmb00}

\begin{figure}[b]
\resizebox{7.5cm}{!}{\includegraphics*[4.5cm,5.5cm][17.5cm,25cm]{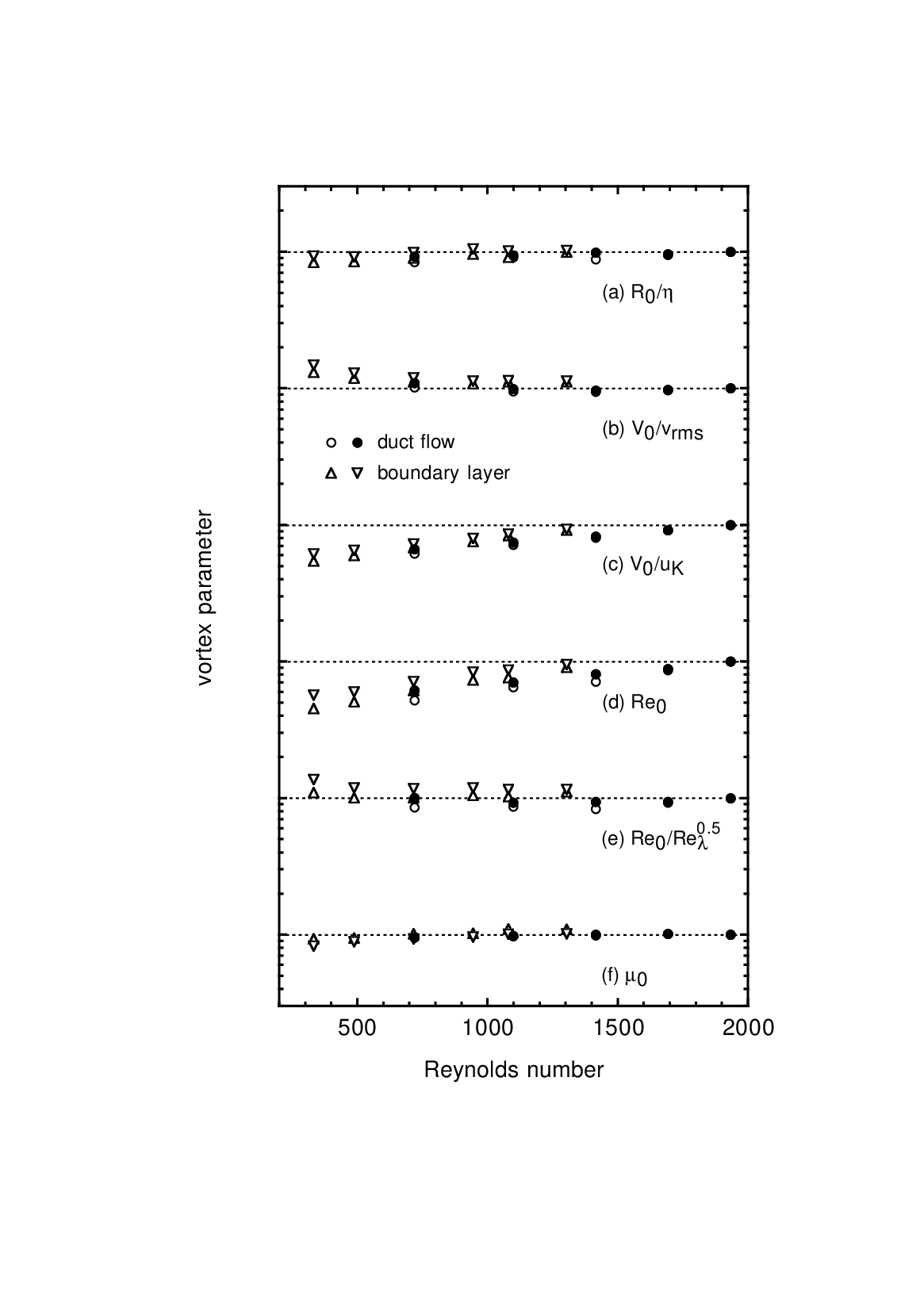}}
\caption{\label{f7} Dependence of parameters for intense vortical structures on Re$_{\lambda}$. (a) $R_0/\eta$. (b) $V_0/\langle v^2 \rangle ^{1/2}$. (c) $V_0/u_K$. (d) Re$_0$. (e) Re$_0$/Re$_{\lambda}^{1/2}$. (f) $\mu_0$. The open and filled circles respectively denote the duct flows for the 0.1\% and 1\% thresholds. The upward and downward triangles respectively denote the boundary layers for the 0.1\% and 1\% thresholds. Each quantity is normalized by its value in the duct flow at Re$_{\lambda} = 1934$ individually for the 0.1\% and 1\% thresholds.}
\end{figure}

\section{Scaling Law}
\label{s6}

Dependence of parameters for intense vortical structures on the microscale Reynolds number Re$_{\lambda}$ and on the configuration for turbulence production, i.e., duct flow or boundary layer, is studied in Fig. \ref{f7}. Each quantity was normalized by its value in the duct flow at Re$_{\lambda} = 1934$ individually for the 0.1\% and 1\% thresholds. That is, we avoid the prefactors that depend on the threshold. When the threshold is high, the radius $R_0$ is small, the maximum circulation velocity $V_0$ is large, and the clustering exponent $\mu_0$ is small as in Table \ref{t2}. We focus on scaling laws of these quantities.

The radius $R_0$ scales with the Kolmogorov length $\eta$ as $R_0 \propto \eta$ [Fig. \ref{f7}(a)]. Thus, intense vortical structures remain to be of smallest scales of turbulence.

The maximum circulation velocity $V_0$ scales with the rms velocity fluctuation $\langle v^2 \rangle ^{1/2}$ as $V_0 \propto \langle v^2 \rangle ^{1/2}$ [Fig. \ref{f7}(b)]. Although the rms velocity fluctuation is a characteristic of the large-scale flow, vortical structures could be formed via shear instability on borders of energy-containing eddies,\cite{cdc95,vsg95,j93} where a small-scale velocity variation could be comparable to the rms velocity fluctuation. The maximum circulation velocity does not scale with the Kolmogorov velocity $u_K$, a characteristic of the small-scale flow, as $V_0 \propto u_K$ [Fig. \ref{f7}(c)].

Direct numerical simulations for intense vortex tubes\cite{j93,jw98} at Re$_{\lambda} \alt 200$ and laboratory experiments for intense vortical structures\cite{b96, mhk04} at Re$_{\lambda} \alt 1300$ derived the scalings $R_0 \propto \eta$ and $V_0 \propto \langle v^2 \rangle ^{1/2}$. We have found that these scalings exist up to Re$_{\lambda} \simeq 2000$, regardless of the configuration for turbulence production.

The scalings of the radius $R_0$ and circulation velocity $V_0$ lead to a scaling of the Reynolds number Re$_0 = R_0 V_0 / \nu$ for the intense vortical structures:\cite{j93,jw98}
%%%%%%%%%%%%%%%%%%%%%%%%%%%%%%%%%%%%%%%%%%%%%%%%%%%%%%%%%%%%%%%%%%%%%%%%%
\begin{subequations}
\label{eq7}
\begin{eqnarray}
\label{eq7a}
&{\rm Re}_0& \propto {\rm Re}_{\lambda}^{1/2} \quad
{\rm if} \
R_0 \propto \eta 
\ {\rm and} \
V_0 \propto \langle v^2 \rangle ^{1/2}, \\
\label{eq7b}
&{\rm Re}_0& = \mbox{constant} \quad
 {\rm if} \
R_0 \propto \eta 
\ {\rm and} \
V_0 \propto u_K.
\end{eqnarray}
\end{subequations}
%%%%%%%%%%%%%%%%%%%%%%%%%%%%%%%%%%%%%%%%%%%%%%%%%%%%%%%%%%%%%%%%%%%%%%%%%
Our result favors the former scaling [Fig. \ref{f7}(e)] rather than the latter [Fig. \ref{f7}(d)]. With an increase of Re$_{\lambda}$, intense vortical structures progressively have higher Re$_0$ and are more unstable.\cite{j93,jw98} Their lifetimes are shorter. It is known\cite{pa01} that the flatness factor $\langle ( \partial _x v )^4 \rangle / \langle ( \partial _x v )^2 \rangle ^2$ scales with Re$_{\lambda}^{0.3}$. Since $\langle ( \partial _x v )^4 \rangle$ is dominated by intense vortical structures, it scales with $\langle v^2 \rangle ^2 / \eta ^4$. Since $\langle ( \partial _x v )^2 \rangle ^2$ is dominated by the background random fluctuation, it scales with $u_K^4 / \eta ^4$. If the number density of intense vortical structures remains the same, we have $\langle ( \partial _x v )^4 \rangle / \langle ( \partial _x v )^2 \rangle ^2 \propto \langle v^2 \rangle ^2 / u_K^4 \propto \mbox{Re}_{\lambda}^2$. The difference from the real scaling implies that vortical structures with $V_0 \simeq \langle v^2 \rangle ^{1/2}$ are less numerous at a higher Reynolds number Re$_{\lambda}$, albeit energetically more important.

The small-scale clustering exponent $\mu_0$ is constant [Fig. \ref{f7}(f)]. A similar result with $\mu_0 \simeq 1$ was obtained from laboratory experiments of the K\'arm\'an flow between two rotating disks\cite{mj05} at Re$_{\lambda} \simeq 400$--1600. The small-scale clustering of intense vortical structures at high Reynolds numbers Re$_{\lambda}$ is independent of the configuration for turbulence production.

Lastly, recall that only intense vortical structures are considered here. For all vortical structures with various intensities, the scalings $V_0 \propto \langle v^2 \rangle ^{1/2}$ and Re$_0 = R_0 V_0 / \nu \propto {\rm Re}_{\lambda}^{1/2}$ are not necessarily expected. For all vortex tubes, in fact, direct numerical simulations\cite{kida,tkmsm04} at Re$_{\lambda} \alt 200$ derived the scaling $V_0 \propto u_K$. The development of an experimental method to study all vortical structures is desirable.

\section{Conclusion}
\label{s7}

The spanwise velocity was measured in duct flows at Re$_{\lambda} = 719$--1934 and in boundary layers at Re$_{\lambda} = 332$--1304 (Table \ref{t1}). We used these velocity data to study features of vortical structures, i.e., vortex tubes and sheets.

We studied the mean velocity profiles of intense vortical structures (Fig. \ref{f4}). The contribution from vortex tubes is dominant. Essentially, our results are those for vortex tubes. The radius $R_0$ is several times the Kolmogorov length $\eta$. The maximum circulation velocity $V_0$ is several tenths of the rms velocity fluctuation $\langle v^2 \rangle ^{1/2}$ and several times the Kolmogorov velocity $u_K$ (Table \ref{t2}). There are the scalings $R_0 \propto \eta$, $V_0 \propto \langle v^2 \rangle ^{1/2}$, and Re$_0 = R_0 V_0 / \nu \propto {\rm Re}_{\lambda}^{1/2}$ (Fig. \ref{f7}).

We also studied the distribution of interval between intense vortical structures. Over large intervals, the distribution obeys an exponential law (Fig. \ref{f5}), which reflects a random and independent distribution of intense vortical structures. Over small intervals, the distribution obeys a power law (Fig. \ref{f6}), which reflects self-similar clustering of intense vortical structures. The clustering exponent is constant, $\mu_0 \simeq 1$ (Table \ref{t2} and Fig. \ref{f7}).

Direct numerical simulations\cite{kida,vm91,j93,jw98,tkmsm04,p94,mtk99} and laboratory experiments\cite{b96,n97,cg99,mhk03,mhk04,mj05} derived some of those features. We have found that they are independent of the Reynolds number and of the configuration for turbulence production, up to Re$_{\lambda} \simeq 2000$ that exceeds the Reynolds numbers of the prior studies.

The Reynolds numbers Re$_{\lambda}$ in our study are still lower than those of  some turbulence, e.g., atmospheric turbulence at Re$_{\lambda} \agt 10^4$. Such turbulence is expected to contain intense vortical structures, because turbulence is more intermittent at a higher Reynolds number Re$_{\lambda}$ and small-scale intermittency is attributable to intense vortical structures. They are expected to have the same features as found in our study. These features appear to have reached asymptotes at Re$_{\lambda} \simeq 2000$ (Fig. \ref{f7}), regardless of the configuration for turbulence production, and hence appear to be universal at high Reynolds numbers Re$_{\lambda}$.

\begin{acknowledgments}
The authors are grateful to T. Gotoh, S. Kida, F. Moisy, M. Takaoka, and Y. Tsuji for interesting discussions.
\end{acknowledgments}

\clearpage


\begin{thebibliography}{999}

\bibitem{f95} U. Frisch, {\it Turbulence, The Legacy of A.N. Kolmogorov} (Cambridge Univ. Press, Cambridge, 1995), Chap. 8.

\bibitem{sa97} K. R. Sreenivasan and R. A. Antonia, ``The phenomenology of small-scale turbulence,'' Annu. Rev. Fluid Mech. {\bf 29,} 435 (1997).

\bibitem{kida} T. Makihara, S. Kida, and H. Miura, ``Automatic tracking of low-pressure vortex,'' J. Phys. Soc. Jpn. {\bf 71,} 1622 (2002). These authors pushed forward the notion that vortex tubes are the elementary structures of turbulence.

\bibitem{vm91} A. Vincent and M. Meneguzzi, ``The spatial structure and statistical properties of homogeneous turbulence,'' J. Fluid Mech. {\bf 225,} 1 (1991).

\bibitem{vm94} A. Vincent and M. Meneguzzi, ``The dynamics of vorticity tubes in homogeneous turbulence,'' J. Fluid Mech. {\bf 258,} 245 (1994).

\bibitem{j93} J. Jim\'enez, A. A. Wray, P. G. Saffman, and R. S. Rogallo, ``The structure of intense vorticity in isotropic turbulence,'' J. Fluid Mech. {\bf 255,} 65 (1993).

\bibitem{jw98} J. Jim\'enez and A. A. Wray, ``On the characteristics of vortex filaments in isotropic turbulence,'' J. Fluid Mech. {\bf 373,} 255 (1998).

\bibitem{tkmsm04} M. Tanahashi, S.-J. Kang, T. Miyamoto, S. Shiokawa, and T. Miyauchi, ``Scaling law of fine scale eddies in turbulent channel flows up to Re$_{\tau} = 800$,'' Int. J. Heat Fluid Flow {\bf 25,} 331 (2004).

\bibitem{p94} A. Pumir, ``Small-scale properties of scalar and velocity differences in three-dimensional turbulence,'' Phys. Fluids {\bf 6,} 3974 (1994).

\bibitem{mtk99} H. Mouri, M. Takaoka, and H. Kubotani, ``Wavelet identification of vortex tubes in a turbulence velocity field,'' Phys. Lett. A {\bf 261,} 82 (1999).

\bibitem{b96} F. Belin, J. Maurer, P. Tabeling, and H. Willaime, ``Observation of intense filaments in fully developed turbulence,'' J. Phys. (Paris) II {\bf 6,} 573 (1996). They studied turbulence velocity fields at Re$_{\lambda} = 151$--5040. We do not consider their results at Re$_{\lambda} \agt 700$, where $\langle ( \partial _x u )^3 \rangle / \langle ( \partial _x u )^2 \rangle ^{3/2}$ and $\langle ( \partial _x u )^4 \rangle / \langle ( \partial _x u )^2 \rangle ^2$ of their data are known to be inconsistent with those from other studies.\cite{sa97} 

\bibitem{n97} A. Noullez, G. Wallace, W. Lempert, R. B. Miles, and U. Frisch, ``Transverse velocity increments in turbulent flow using the RELIEF technique,'' J. Fluid Mech. {\bf 339,} 287 (1997).

\bibitem{cg99} R. Camussi and G. Guj, ``Experimental analysis of intermittent coherent structures in the near field of a high Re turbulent jet flow,'' Phys. Fluids {\bf 11,} 423 (1999).

\bibitem{mhk03} H. Mouri, A. Hori, and Y. Kawashima, ``Vortex tubes in velocity fields of laboratory isotropic turbulence: dependence on the Reynolds number,'' Phys. Rev. E {\bf 67,} 016305 (2003).

\bibitem{mhk04} H. Mouri, A. Hori, and Y. Kawashima, ``Vortex tubes in turbulence velocity fields at Reynolds numbers Re$_{\lambda} \simeq 300$--1300,'' Phys. Rev. E {\bf 70,} 066305 (2004).

\bibitem{sd98} K. R. Sreenivasan and B. Dhruva, ``Is there scaling in high-Reynolds-number turbulence?,'' Prog. Theor. Phys. Suppl. {\bf 130,} 103 (1998).

\bibitem{note0} The two wires individually respond to all the $u$, $v$, and $w$ components. Since the measured $u$ component corresponds to the sum of the responses of the two wires, it is contaminated with the $w$ component. Since the measured $v$ component corresponds to the difference of the responses, it is free from the $w$ component.

\bibitem{l82} T. S. Lundgren, ``Strained spiral vortex model for turbulent fine structure,'' Phys. Fluids {\bf 25,} 2193 (1982).

\bibitem{note2} For convenience, when consecutive increments are all above the threshold, each increment is taken to determine the center of a vortex. This is somewhat unreasonable but does not cause serious problems, judging from Fig. \ref{f2} where mean velocity profiles were obtained practically in the same manner.

\bibitem{note1} While the experimental curves in Figs. \ref{f3} and \ref{f4} are mere loci of discrete data points, we applied smoothing to the tails of the experimental curves in Figs. \ref{f5} and \ref{f6}.

\bibitem{mj04} F. Moisy and J. Jim\'enez, ``Geometry and clustering of intense structures in isotropic turbulence,'' J. Fluid Mech. {\bf 513,} 111 (2004).

\bibitem{mj05} F. Moisy and J. Jim\'enez, ``Clustering of intense structures in isotropic turbulence: numerical and experimental evidence,'' in {\it IUTAM Symposium on Elementary Vortices and Coherent Structures: Significance in Turbulence Dynamics}, edited by S. Kida (Springer, Dordrecht, 2006), p. 3.

\bibitem{sm05} K. Sassa and H. Makita, ``Reynolds number dependence of elementary vortices in turbulence,'' in {\it Engineering Turbulence Modelling and Experiments 6}, edited by W. Rodi and M. Mulas (Elsevier, Oxford, 2005), p. 431.

\bibitem{note3} The positive excursion might be partially induced by fluctuation of the instantaneous velocity $U+u$ at which a structure passes the probe. Under Taylor's frozen-eddy hypothesis, the velocity increment over the sampling interval $U/f_s$ is more intense for a faster-moving structure, which is more likely to be incorporated in our conditional averaging.\cite{mhk03} Other mechanisms might be also at work.

\bibitem{t92} M. Kholmyansky, A. Tsinober, and S. Yorish, ``Velocity derivatives in the atmospheric surface layer at Re$_{\lambda} = 10^4$,'' Phys. Fluids {\bf 13,} 311 (2001).

\bibitem{affl94} P. Abry, S. Fauve, P. Flandrin, and C. Laroche, ``Analysis of pressure fluctuations in swirling turbulent flows,'' J. Phys. (Paris) II {\bf 4,} 725 (1994).

\bibitem{cdc95} O. Cadot, S. Douady, and Y. Couder, ``Characterization of the low-pressure filaments in a three-dimensional turbulent shear flow,'' Phys. Fluids {\bf 7,} 630 (1995).

\bibitem{vsg95} E. Villermaux, B. Sixou, and Y. Gagne, ``Intense vortical structures in grid-generated turbulence,'' Phys. Fluids {\bf 7,} 2008 (1995).

\bibitem{lvmb00} A. La Porta, G. A. Voth, F. Moisy, and E. Bodenschatz, ``Using cavitation to measure statistics of low-pressure events in large-Reynolds-number turbulence,'' Phys. Fluids {\bf 12,} 1485 (2000).

\bibitem{pa01} B. R. Pearson and R. A. Antonia, ``Reynolds-number dependence of turbulent velocity and pressure increments,'' J. Fluid Mech. {\bf 444,} 343 (2001).


\end{thebibliography}
\end{document}